\documentclass[reprint,aps,ams,amsmath,prx,twocolumn,superscriptaddress,longbibliography]{revtex4-1}
\usepackage{graphics}
\usepackage{graphicx}
\usepackage{epsfig}
\usepackage{amsmath}
\usepackage{amssymb}
\usepackage{amsfonts}
\usepackage{bm}
\usepackage{mathrsfs}
\usepackage{color}
\usepackage{bbm}
\usepackage{hyperref}
\usepackage[normalem]{ulem}
\usepackage{comment}
\usepackage{soul}

\begin{document}

\title{Dragging of electric current by hydrodynamic flow  at charge neutrality}

\author{Dmitry Zverevich}
\affiliation{Department of Physics, University of Wisconsin-Madison, Madison, Wisconsin 53706, USA}

\author{Alex Levchenko}
\affiliation{Department of Physics, University of Wisconsin-Madison, Madison, Wisconsin 53706, USA}

\author{A. V. Andreev}
\affiliation{Department of Physics, University of Washington, Seattle, Washington 98195, USA}

\date{October 31, 2025}

\begin{abstract}
We develop a theory of drag in graphene double layers near charge neutrality. We work in the regime of electron hydrodynamics and account for interlayer correlations of charge puddle disorder. The drag resistivity is expressed in terms of the viscosity, intrinsic conductivity of the electron liquid, and the correlation function of the puddle disorder. 
The contributions of the interlayer transfer of momentum and energy to drag have opposite signs. This leads to a nonmonotonic dependence of  the drag resistivity on the carrier density. For layer-symmetric doping, the drag resistivity changes sign as a function of the carrier density.  At interlayer separations shorter than the disorder correlation length, the transconductivity  
saturates to the disorder-induced enhancement of the intralayer conductivity. 
We provide quantitative estimates of the effect for Dirac electron liquids in monolayer graphene and bilayer graphene double-layer devices.
\end{abstract}

\maketitle


Quantum devices comprised of spatially separated conducting electron systems enable exploring electron correlation at the mesoscale. Examples of fascinating phenomena driven by correlations and quantum or thermal fluctuations in electron double layers (EDL) include the Casimir effect \cite{Casimir1948,CasimirPolder1948}, van der Waals forces \cite{Lifshitz1956,Dzyaloshinskii1960}, radiative near field heat transfer \cite{Polder1971,Pendry1999}, mutual friction \cite{Pogrebinskii1977,Price1983,Levitov1989}, excitonic superfluidity \cite{LozovikYudson,Eisenstein2004}, and the physics of quantum Hall fluids \cite{Rezayi,WenZee}.  

In EDL systems, the electron-electron correlations can be probed by Coulomb drag measurements \cite{Rojo1999,Eisenstein2005,RMP-Drag}. During the past decade, the experimental \cite{Tutuc2011,Tutuc2012,Ponomarenko2012,Titov2013,Dean2016,Chen2017,Das2020,Zhu2020,Kim2021,Bandurin2022,Zhu2023,Ponomarenko2024,Wang2024} and theoretical \cite{TseHu2007,Katsnelson2011,Hwang2011,CastroNeto2011,Song2012,Ostrovsky2012,Carrega2012,Amorim2012,Lux2012,Lux2013,Schutt2013,Abanin2013,Xie2017,Adam2018,Chudnovskiy2024} efforts devoted to the exploration of Coulomb drag phenomena have been centered around monolayer graphene (MLG) and bilayer graphene (BLG) double-layer devices. In contrast to two-dimensional semiconductor quantum well heterostructures \cite{Gramila1991,Gramila1993,Lilly1998,Pillarisetty2002,Kellogg2003,Pillarisetty2003}, these systems offer substantial advantages, which are rooted in their unprecedented degree of tunability. (i) Independent gate control allows drag measurement between carriers of the same type, e.g. electron-electron ($ee$) and hole-hole ($hh$) EDLs, or alternatively between opposite charges of electron-hole ($eh$) liquids. (ii) The system can be probed in a broad range of particle densities $n$ from global charge neutrality to the carrier concentration of the order of $\sim 10^{12}$ cm$^{-2}$ in each layer. (iii) This enables access to both the quantum degenerate regime, $T<E_F$, and 
the  classical regime, $T>E_F$, where $E_F$ is the Fermi energy,
allowing to tune the ratio between the interlayer spacing to the inelastic intralayer mean free path.
(iv) Twist angle between the layers provides an additional degree of freedom to explore emergent quantum friction phenomena.   

The observed drag response in graphene~\cite{Ponomarenko2012,Dean2016} revealed a number of anomalous features. The most interesting aspect of the experimental data is the nonzero drag resistivity observed at the double neutrality point (DNP). The drag is positive, and its temperature dependence is nonmonotonic: it initially increases with increasing temperature and subsequently diminishes as the temperature is raised further (typically above $\sim 100$ K), often developing a sharp peak. Away from charge neutrality, the drag resistivity changes sign for bilayers at matched densities of the same carrier type.

Near charge neutrality, the correlations of the electron liquid in graphene become especially pronounced. At this point, the Fermi-liquid does not apply, and the electron-electron interactions  in both MLG  and BLG become strong. This makes the construction of a quantum theory of the electron liquid a challenging problem. Another consequence of strong correlations is that at finite temperatures, the relaxation rate due to \emph{ee} collisions becomes very short. For example, in MLG devices with the massless Dirac spectrum, the dimensionless interaction constant is $e^2/\hbar v_F \approx 2.2$, and \emph{ee} relaxation rate becomes of the order of the temperature $T$, reaching the Planckian bound. As a result, the hydrodynamic description of such strongly correlated liquids becomes applicable from very short distances, of the order of the thermal de Broglie length~\cite{Spivak2010,AKS}. 

The salient feature of electron hydrodynamics at charge neutrality is decoupling of electric current from the hydrodynamic flow, which corresponds to the pure flow of heat/entropy.  As a consequence,  the fluctuation-driven transfer of momentum and energy between the layers produces dragging of the hydrodynamic flow of heat but does not affect the charge current. 

Here, we develop a hydrodynamic theory of electric current drag near charge neutrality. We show that it is caused mainly by the entrainment of charge by the hydrodynamic flow, which is induced by the long-range charge puddle disorder. 
The latter is believed to exist even in the cleanest graphene devices \cite{Crommie2009,LeRoy2011}, which are ideal candidates for the realization of hydrodynamic electron transport \cite{Crossno2016,Ghahari2016}. 

\begin{figure}[t!]
\includegraphics[width=\linewidth]{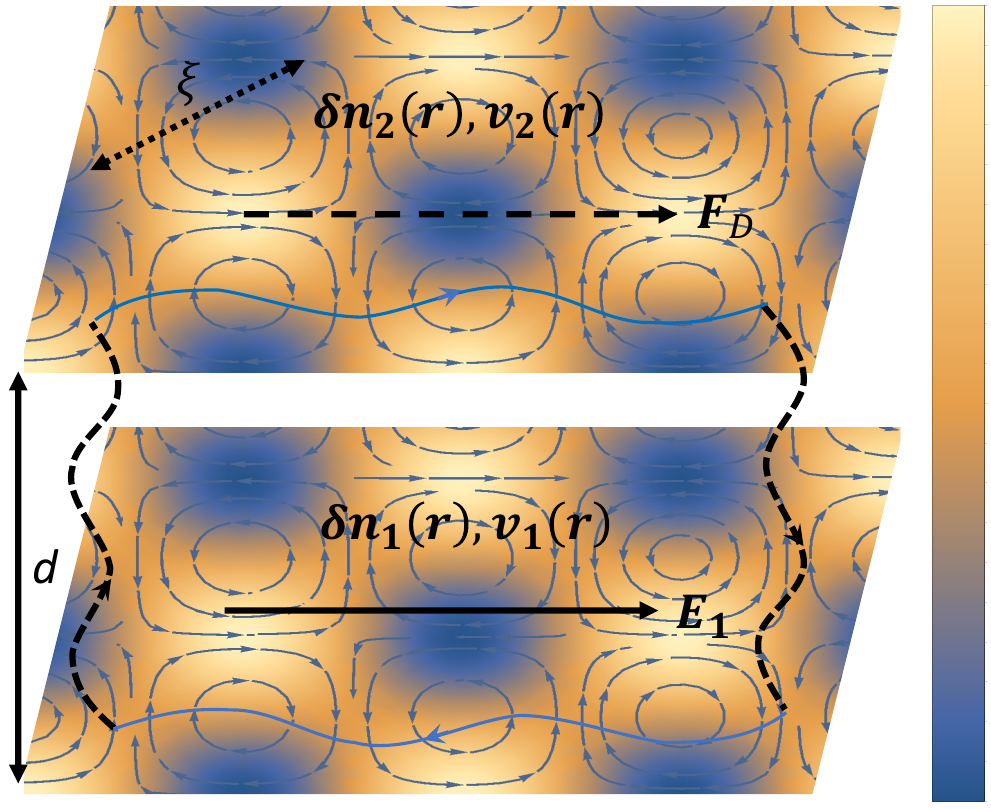}
\caption{Setup for the EDL device with interlayer separation $d$. The color map in each layer represents electron-hole puddles modeled for a checkerboard potential with the correlation radius $\xi$. The positive/negative densities, $\delta n(\bm{r})$, are shown in yellow/blue, respectively. The driving electric field $\bm{E}_1$ induces temperature modulations and a vortical flow
in the active layer. The transfer of momentum induces a co-moving vortical flow in the passive later as shown by the closed stream lines. The interlayer heat transfer caused by temperature modulations is indicated by broken dashed lines. It drives counter-propagating longituninal flows in the two layers, shown by open blue lines. }
\label{fig:EDL}
\end{figure}

We consider an electronic double layer that consists of two copies of two-dimensional electron systems separated by the distance $d$, see Fig. \ref{fig:EDL}. We assume that the layers are subjected to a disorder potential $U(\bm{r})$ whose characteristic spatial scale $\xi $ exceeds the  $(ee)$ mean free path $l$. In this regime electron transport may be described using the hydrodynamic approach~\cite{NGMS,Lucas-Fong,Narozhny2022}. In the presence of electron-hole puddles, the electric current and hydrodynamic flow become coupled even in samples that are on average charge-neutral. The electric current in the active layer (layer 1) induces in it a hydrodynamic flow, in which the local temperature modulations (that are linear in the current) and hydrodynamic velocity are  correlated with the disorder potential. This flow, in turn will induce a transfer of energy and momentum to the passive layer (layer 2), as illustrated in Fig.~\ref{fig:EDL}. In the regime $d \ll \xi$ the
rates of the interlayer momentum transfer, $\dot{\bm{P}}_{12} (\bm{r})$, and energy transfer, $\dot{E}_{12}(\bm{r})$, are proportional to the local interlayer difference in temperature and the hydrodynamic velocity, 
\begin{subequations}
\label{eq:interlayer_currents}
\begin{align}
    \label{eq:local_rates}
    &\dot{\bm{P}}_{12} (\bm{r}) = \, k_D (\bm{v}_1(\bm{r}) - \bm{v}_2(\bm{r})), \\
    &\dot{E}_{12}(\bm{r}) =\varkappa_T (T_1(\bm{r}) - T_2(\bm{r})).  
\end{align}
\end{subequations}
Here we introduced  the interlayer friction coefficient $k_D$ and interlayer thermal conductivity $\varkappa_T$ as phenomenological parameters.   
In hydrodynamic theory they can be computed analytically in the limit of large interlayer separation $d>\lambda_T$, see Refs. \cite{Levchenko2022,Zverevich2023}.

The interlayer energy and momentum currents \eqref{eq:interlayer_currents} will induce a spatially-inhomogeneous distribution of temperature and hydrodynamic velocity in the passive layer. Because of the correlations of disorder in the passive and active layers, this will produce a uniform charge current in the passive layer.
The importance of disorder correlations for electrical drag was emphasized in Refs.~\cite{Gornyi1999,Schutt2013,Song2012}. The momentum drag ($P$-drag) contribution is proportional to the coefficient $k_D$ in Eq.~\eqref{eq:local_rates}. In pristine systems at double charge neutrality the $P$-drag contribution vanishes because electric current is decoupled  from the hydrodynamic flow.  
To describe drag at charge neutrality, the energy-driven drag mechanism  ($E$-drag) was proposed in Ref. \cite{Song2012}, in which electrical drag was caused by the correlations between the local interlayer heat transfer rate and thermoelectric properties in the electron liquid. 
In the hydrodynamic entrainment mechanism discussed below, the contribution of energy transfer to the electric drag is caused by thermally-induced pressure gradients in the passive layer. They drive an inhomogeneous hydrodynamic flow correlated with the charge puddle disorder, creating a macroscopic charge current. This produces much stronger $E$-drag than the thermoelectric mechanism. Furthermore, as we show below, in the presence of the puddle disorder, $P$-drag does not vanish at double charge neutrality, and produces the dominant contribution to electric drag. This results in a sign change of the overall drag coefficient as a function of the average doping in the two layers. This sign change was observed experimentally Refs.~\cite{Ponomarenko2012,Dean2016}.

We work near charge neutrality, which we define as the regime where the local charge density (in units of electron charge), $n_i(\bm{r})$, is small compared to the entropy density $s (\bm{r})$. 
In systems without Galilean invariance, in addition to the convective contribution proportional to the flow velocity, $ e n_i \bm{v}_i$,
the electric current $\bm{j}^e_i$ receives contributions from the intrinsic conductivity of the electron fluid, $\sigma_0$, and intrinsic thermoelectric conductivity, $\gamma_0$, 
\begin{align}
    \label{eq:cont-n}
    \bm{\nabla}\cdot\bm{j}^e_i=0,\quad 
    \bm{j}^e_i = & \, e n_i \bm{v}_i + \sigma_0 \bm{\mathcal{E}}_i - \frac{e \gamma_0}{T}\bm{\nabla} T_i,
\end{align}
where $\bm{\mathcal{E}}$ is the local electromotive force (EMF). The current in each layer labeled with index $i=1,2$ is conserved independently in the absence of interlayer tunneling of electrons.

In contrast, the momentum and energy evolution equations are affected by interlayer transfer in Eq.~\eqref{eq:interlayer_currents}. In linear response, the energy evolution equation reduces to the continuity equation for the entropy current density,
\begin{equation}\label{eq:cont-s}
\bm{\nabla}\cdot\bm{j}^s_i=-\frac{\dot{E}_{ij}}{T},\quad \bm{j}^{s}_i=s_i\bm{v}_i+\frac{\gamma_0}{T}e\bm{\mathcal{E}}_i-\frac{\kappa_0}{T}\nabla T_i
\end{equation}
where $\kappa_0$ is the intrinsic thermal conductivity of the fluid. Note that, when working in linear response, we may neglect an entropy production term due to
electron collisions. The evolution of momentum density can be described as 
a local force-balance condition that takes the form of the linearized Navier-Stokes equation 
\begin{equation}\label{eq:NS}
\eta\bm{\nabla}^2\bm{v}_i-s_0\bm{\nabla}T_i+e n_i\bm{\mathcal{E}}_i=\dot{\bm{P}}_{ij}.
\end{equation}
For simplicity, we neglected terms with the bulk viscosity, $\bm{\nabla}[\zeta(\bm{\nabla}\cdot\delta\bm{v}_i)]$, which is justified as $\zeta$ is known to vanish in systems with parabolic and linear spectrum \cite{LL-V10}. In Eq. \eqref{eq:NS} the pressure gradient was absorbed into the EMF $e \bm{\mathcal{E}}$, which is equal to the gradient of the electrochemical potential, using the thermodynamic identity $dP = n d\mu +s dT$. 
The puddle disorder enters the governing equations via the inhomogeneous component of the equilibrium density $n_i(\bm{r})=n_0+\delta n_i(\bm{r})$. 

The problem of determining the drag conductivity 
$\sigma_D$ in a bilayer system reduces to solving the coupled linear flow equations and calculating the net charge flux in the passive layer, 
$\bm{j}^e_2$, in response to the applied field 
$\bm{E}_1$ in the active layer, such that $\bm{j}^e_2=\sigma_D\bm{E}_1$.
To solve the system of governing equations, we separate various quantities into spatially uniform and nonuniform components. 

We first determine the electron flow in both layers at a dual (on average) charge neutrality $n_{0}\to0$ (note that $\gamma_0\to0$ in this limit). For this purpose it is convenient  
to separate the velocity into the longitudinal, $\bm{v}^l_i$, and the transverse, $\bm{v}^t_i$, components, so that 
\begin{equation}
\bm{v}_i=\bm{v}^l_i+\bm{v}^t_i.
\end{equation} 
Since Eqs.~\eqref{eq:cont-n} and \eqref{eq:cont-s} do not involve $\bm{v}^t_i$, 
the latter determined from the transverse part of Eq.~\eqref{eq:NS}. The spatial modulations of EMF, temperature and $\bm{v}^l_i$, are determined from the longitudinal part of \eqref{eq:NS} and Eqs.~\eqref{eq:cont-n} and \eqref{eq:cont-s}. 
The result for the Fourier components of the inhomogeneous part of the hydrodynamic velocity is found in the form \cite{SM}
\begin{subequations} \label{eq:v-t}
\begin{align}
\delta\bm{v}^t_1(\bm{q})=\frac{\eta q^2+k_D}{\eta q^2(\eta q^2+2k_D)}\left[\bm{E}_1-\frac{\bm{q}(\bm{q}\cdot\bm{E}_1)}{q^2}\right]e\delta n_1(\bm{q}), \\
\delta\bm{v}^t_2(\bm{q})=\frac{k_D}{\eta q^2(\eta q^2+2k_D)}\left[\bm{E}_1-\frac{\bm{q}(\bm{q}\cdot\bm{E}_1)}{q^2}\right]e\delta n_1(\bm{q}). 
\end{align}
\end{subequations}
The solution for the longitudinal components also follows from Eq. \eqref{eq:NS}, however, it is more involved as one has to invoke continuity equations \eqref{eq:cont-n}--\eqref{eq:cont-s} 
to express local EMF and temperature gradients. After somewhat lengthy but otherwise straightforward calculation we find 
\begin{subequations}\label{eq:v-l} 
\begin{align}
\delta\bm{v}^l_1(\bm{q})=\frac{A(\varkappa_T+\kappa_0q^2)-\varkappa_TB}{A^2-B^2}e\delta n_1(\bm{q})\frac{\bm{q}(\bm{q}\cdot\bm{E}_1)}{q^2}, \\
\delta\bm{v}^l_2(\bm{q})=\frac{B(\varkappa_T+\kappa_0q^2)-\varkappa_TA}{A^2-B^2}e\delta n_1(\bm{q})\frac{\bm{q}(\bm{q}\cdot\bm{E}_1)}{q^2},
\end{align}
\end{subequations}
where we have introduced for compactness 
\begin{subequations}
\begin{align}
&A=Ts^2_0q^2+(\eta q^2+k_D)(\varkappa_T+\kappa_0q^2)+\varkappa_Tk_D,\\
&B=(\varkappa_T+\kappa_0q^2)k_D+(\eta q^2+k_D)\varkappa_T. 
\end{align}   
\end{subequations}

These results enable us to obtain the current induced by hydrodynamic entrainment. We notice from Eqs. \eqref{eq:v-t} and \eqref{eq:v-l} that $P$-drag primarily affects the transverse component of the hydrodynamic velocity, whereas the $E$-drag affects the longitudinal component. Therefore, below we identify the contributions of the transverse and longitudinal velocities to the current with $P$- and $E$-drag, respectively, $\bm{j}^e_2=\bm{j}_P+\bm{j}_E$, 
where $\bm{j}_P=\langle e\delta n_2\delta\bm{v}^t_2\rangle=\sigma_P\bm{E}_1$ and $\bm{j}_E=\langle e\delta n_2\delta\bm{v}^l_2\rangle=\sigma_E\bm{E}_1$. Here $\langle\ldots\rangle=\int(\ldots)d\bm{r}/S$ denotes spatial average over the 2D system with the surface area $S$. For the respective conductivities we find as a result 
\begin{subequations}
\label{eq:sigmaD}
\begin{equation}
\sigma_D=\sigma_P+\sigma_E, 
\end{equation}
\begin{equation}\label{eq:sigmaDP}
\sigma_P=\frac{e^2}{2}\int_{\bm{q}}\frac{k_D}{\eta q^2+2k_D}\frac{\langle\delta n_1(\bm{q})\delta n_2(-\bm{q})\rangle}{\eta q^2}, 
\end{equation}
\begin{equation}\label{eq:sigmaDE}
\sigma_E=-\frac{e^2}{2}\int_{\bm{q}}\frac{\varkappa_T}{Ts^2_0q^2+4\varkappa_Tk_D}\langle\delta n_1(\bm{q})\delta n_2(-\bm{q})\rangle.\\
\end{equation}
\end{subequations}
For simplicity, Eq. \eqref{eq:sigmaDE} is written to the leading order in $Ts^2_0\gg\{\varkappa_T\eta,\kappa_0k_D\}$ (see \cite{SM} for the complete expression).

Equation \eqref{eq:sigmaD} for the drag conductivity at dual charge neutrality applies to both MLG and BLG devices.  The $E$-drag contribution, $\sigma_E$, is of the opposite sign compared to the $P$-drag term. This can be understood as follows. Charge puddle disorder induces temperature gradients in the active layer, which are linear in the driving current. The ensuing interlayer heat transfer drives circulating hydrodynamic flow that has opposite direction in the two layers, as indicated by the loop involving interlayer heat transfer in Fig.~\ref{fig:EDL}. Because of the correlated puddle disorder, advection of charge by this flow gives a negative contribution to drag conductivity. 
In contrast, $P$-drag induces in the passive layer a vortical flow that is correlated with that in the active layer, as shown by the blue streamlines in Fig. 1. This produced a positive contribution to drag conductivity. The $P$-drag contribution exceeds the $E$-drag contribution at double charge neutrality for arbitrary correlation function of charge puddle disorder. 

We note that for the long-range disorder, when $k_D\xi^2\gg\eta$, the coefficient of the drag friction drops out from Eq. \eqref{eq:sigmaDP}, since $k_D/(\eta q^2+2k_D)\to1/2$, so that $\sigma_P$ is expressed solely in terms of the viscosity and correlation function of density fluctuations. In this regime $P$-drag conductivity is independent of the interlayer spacing.  Moreover, for perfect interlayer correlations of the disorder potential, the $P$-drag conductivity becomes equal to half of the disorder-induced enhancement of the intralayer conductivity (see Ref. \cite{Li2020}). 

Evaluation of the drag conductivities for both $P$ and $E$ mechanisms can be readily generalized to the regime of nonzero layer densities by following the general scheme developed in Refs.~\cite{Li2020,Lucas2016}. We restrict our consideration to the cases of symmetric, $n_1=n_2=n_0$, and antisymmetric, $n_1=-n_2=n_0$, doping of layers. 
Working in leading order in $n_0\ll s_0$, we note that thermoelectric current is weak near charge neutrality since $\gamma_0/T\propto (n_0/s_0)\ll1$. For this reason, we neglect the contribution to the current arising from the intrinsic thermoelectric effect at small density. Deferring the technical details of the calculations to the Supplemental Material~\cite{SM}, below we present the main results. 
The drag conductivity is given by 
\begin{equation}
\sigma_D(n_0)\approx\sigma_P\pm\frac{e^2k_Dn^2_0}{k(k+2k_D)},
\end{equation}  
where $\sigma_P$ comes from Eq. \eqref{eq:sigmaDP}. The plus sign corresponds to the symmetric ($ee$ or $hh$) doping, whereas the minus sign describes antisymmetric ($eh$) doping. In the expression above $k$ is the intralayer friction coefficient~\cite{Li2020,Lucas2016}, which relates the disorder-induced friction force to the spatial average of hydrodynamic velocity, $\bm{F} = - k \bm{v}$ \cite{SM}.  In the same limit, the intralayer conductivity is given by \cite{Li2020} 
\begin{equation}
\sigma(n_0)\approx\sigma_0\left[1+\chi_\sigma+\frac{e^2}{\sigma_0}\frac{n^2_0}{k}\right],
\end{equation}
where we introduced 
\begin{equation}
\chi_\sigma=\frac{e^2}{2\sigma_0}\int_{\bm{q}}\frac{\langle|\delta n(\bm{q})|^2\rangle}{\eta q^2}, \quad k=\frac{e^2}{2\sigma_0}\int_{\bm{q}}\langle |\delta n(\bm{q})|^2\rangle.  
\end{equation}
The term in $\sigma(n_0)$ with $\chi_\sigma$ is the disorder-induced renormalization of the intrinsic conductivity at charge neutrality. The last term defines an additional contribution at finite density (convective contribution $en_0\bm{v}$) that is stabilized by the emergent disorder-induced intralayer friction $\bm{v}=en_0\bm{E}_1/k$. 

\begin{figure}[t!]
\includegraphics[width=\linewidth]{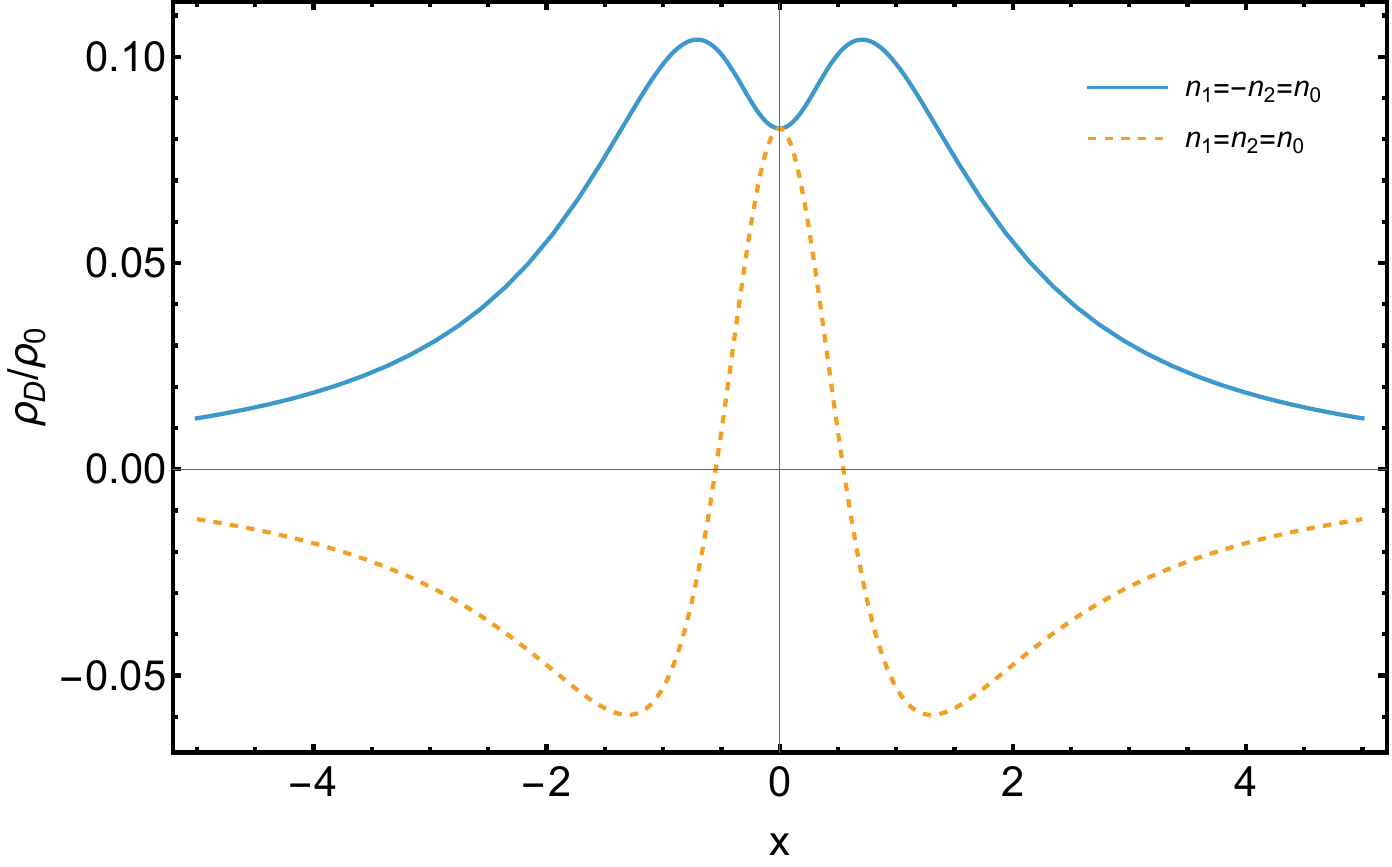}
\caption{Density dependence of the normalized drag resistivity of a bilayer system for the matched electron-electron and electron-hole densities subjected to a perfectly correlated long-range disorder.  
The value of the drag resistivity at dual neutrality point is sent by the disorder-induced enhancement of conductivity characterized by the parameter $\chi_\sigma$ (on the plot $\chi_{\sigma}=0.1$ and $p=1/3$). 
 The density is plotted in the dimensionless units of $x=en_{0}/\sqrt{k\sigma_{0}}$.}
\label{fig:rhoD}
\end{figure}

Inverting the transconductivity matrix to obtain drag resistivity, and taking also the limit of perfect interlayer disorder correlation, one finds
\begin{equation}\label{eq:rhoD}
\rho_D(n_0)\approx-\frac{\sigma_D(n_0)}{\sigma^2(n_0)}=\sigma^{-1}_0\frac{\chi_\sigma\mp p x^2}{(1+\chi_\sigma+x^2)^2}.
\end{equation}     
The dimensionless density is introduced here in units of the variable $x=\sqrt{\frac{e^2}{\sigma_0}}\frac{n_0}{\sqrt{k}}$ and $p=k_D/(k+2k_D)$. We plot $\rho_D(n_0)$ in Fig. \ref{fig:rhoD} for both cases of $ee$-type and $eh$-type bilayers. We find a change of sign of the drag resistivity at finite density, which is qualitatively consistent with the experiment \cite{Ponomarenko2012}. The sign change is driven by $P$-drag and can be understood as follows. The flow induced by $P$ drag in the passive layer is codirected with that in the active layer. For sufficiently large doping, this produces a negative contribution to drag.

Drag resistivity near charge neutrality is strongly temperature dependent, as can be seen from the linear screening theory. In the hydrodynamic regime, the Thomas-Fermi screening radius, $r_{\text{TF}}=1/(2\pi e^2\nu)$,  where $\nu=\partial n/\partial\mu$ is the thermodynamic density of states, is short compared to $\xi$. As a consequence, at length scales of the order of $\xi$, the compressibility of the electron liquid is dominated by the Coulomb interaction,
\begin{equation}
\delta n_{1,2}(\bm{q})/q=-U(\bm{q})/(2\pi e^2).
\end{equation} 
Substituting this into Eq.~\eqref{eq:sigmaDP} we get the estimate for the drag resistivity 
\begin{equation}\label{eq:rhoD}
\rho_D(n_0\to0)=\frac{e^2}{\sigma^2_0\eta}\frac{\langle U^2\rangle}{(2\pi e^2)^2}.
\end{equation} 
Estimating the viscosity in MLG as $\eta \sim (T/v)^2$ \cite{MuellerFritz2009}, with $v$ being the band velocity, we get
\begin{equation}\label{eq:rhoD-estimate}
\frac{\rho_D}{\rho_Q}\sim \frac{1}{\alpha^2_g}\left(\frac{\sigma_Q}{\sigma_0}\right)^2\frac{\langle U^2\rangle}{T^2}, \quad T>v/\xi,
\end{equation}  
where $\rho_Q=\sigma^{-1}_Q=h/e^2$ is the quantum of resistance. 
Using the measured values of the intrinsic conductivity $\sigma_0$, and the typical magnitude of the disorder fluctuations $U\sim 5$ meV extracted from the scanning tunneling probes, at $T\sim100$ K and $\alpha_g\sim1$, we get $\rho_D\lesssim 50 \Omega$. 
The power law decay of the drag resistivity with an increase of temperature predicted by Eq. \eqref{eq:rhoD-estimate} is qualitatively consistent with the observed behavior in the experiment, see inset in the Fig. 3(a) of Ref. \cite{Ponomarenko2012}. The role of correlated macroscopic inhomogeneities in the hydrodynamic limit of drag was discussed in Ref. \cite{Schutt2013} [see their Appendix Sec. 4-B] and the conclusion of $\rho_D\propto 1/T^2$ was also reached at DNP albeit within the different model considerations and limiting cases. 

A similar estimation can be made for the BLG devices. Even though viscosity has not been calculated analytically, from the viscosity-to-entropy bound conjecture of the strong-coupling theory \cite{KSS2005}, one can infer that $\eta\sim m^*T$, where $m^*$ is the effective mass of the band structure. Therefore, based on Eq. \eqref{eq:rhoD} we expect drag resistance to diminish as $\rho_D\propto 1/T$.    

In summary, we developed a theory of drag in graphene double layers near charge neutrality. Drag is caused by the advection of charge by the hydrodynamic flow, which is correlated with the charge puddle disorder and is driven by both $P$- and $E$-drag. The $P$-drag provides the dominant contribution and leads to a sign change of the drag conductivity at finite doping, as observed in recent experiments in MLG and BLG devices. 

We thank I. Gornyi, L. Levitov, and B. Narozhny for constructive discussions.
The work of D. Z. and A. L. was supported by NSF Grant No. DMR-2452658 and H. I. Romnes Faculty Fellowship provided by the University of Wisconsin-Madison Office of the Vice Chancellor
for Research and Graduate Education with funding from the Wisconsin Alumni Research Foundation. The work of A. V. A. was
supported by the National Science Foundation (NSF) Grant No. DMR-2424364.

\bibliography{biblio}
\end{document}